\documentclass[runningheads]{llncs}
\usepackage{graphicx}
\usepackage{cite}
\usepackage{amsmath,amssymb,amsfonts}
\usepackage{algorithmic}
\usepackage{graphicx}
\usepackage{textcomp}
\usepackage{adjustbox}
\usepackage{multirow}
\usepackage{bbold}
\usepackage{siunitx}
\usepackage{hyperref}       
\usepackage{xcolor}
\usepackage{sidecap}

\usepackage{booktabs}
\usepackage[mathscr]{eucal}

\hypersetup{
    colorlinks,
    linkcolor={blue!50!black},
    citecolor={blue!50!black},
    urlcolor={blue!50!black}
}
\usepackage{amsmath, bm}
\usepackage{hyperref}

\usepackage{pifont}
\newcommand{\xmark}{\ding{53}}%

\usepackage{floatrow}
\newfloatcommand{capbtabbox}{table}[][\FBwidth]

\begin{document}

\title{EndoDAC: Efficient Adapting Foundation Model for Self-Supervised Depth Estimation from Any Endoscopic Camera
}
\author{Beilei Cui\inst{*,1} \and
Mobarakol Islam\inst{*,2} \and
Long Bai\inst{*,1} \and
An Wang\inst{1} \and
Hongliang Ren\inst{\dag,1,3}}

\authorrunning{Cui et al.}
%

\institute{Dept. of Electronic Engineering, The Chinese University of Hong Kong, Hong Kong SAR, China \and
Wellcome/EPSRC Centre for Interventional and Surgical Sciences (WEISS), University College London, UK \and 
Dept. of Biomedical Engineering, National University of Singapore, Singapore \\
\email{beileicui@link.cuhk.edu.hk, mobarakol.islam@ucl.ac.uk, b.long@link.cuhk.edu.hk, wa09@link.cuhk.edu.hk, hlren@ee.cuhk.edu.hk}}

\maketitle              
\begin{abstract}

Depth estimation plays a crucial role in various tasks within endoscopic surgery, including navigation, surface reconstruction, and augmented reality visualization. Despite the significant achievements of foundation models in vision tasks, including depth estimation, their direct application to the medical domain often results in suboptimal performance. This highlights the need for efficient adaptation methods to adapt these models to endoscopic depth estimation. We propose Endoscopic Depth Any Camera (EndoDAC) which is an efficient self-supervised depth estimation framework that adapts foundation models to endoscopic scenes. Specifically, we develop the Dynamic Vector-Based Low-Rank Adaptation (DV-LoRA) and employ Convolutional Neck blocks to tailor the foundational model to the surgical domain, utilizing remarkably few trainable parameters. Given that camera information is not always accessible, we also introduce a self-supervised adaptation strategy that estimates camera intrinsics using the pose encoder. Our framework is capable of being trained solely on monocular surgical videos from any camera, ensuring minimal training costs. Experiments demonstrate that our approach obtains superior performance even with fewer training epochs and unaware of the ground truth camera intrinsics. Code is available at~\url{https://github.com/BeileiCui/EndoDAC}.


\end{abstract}

\renewcommand{\thefootnote}{}
\footnotetext{\inst{*} Authors contributed equally to this work.}
\footnotetext{\inst{\dag} Corresponding Author.}

\section{Introduction}

Depth estimation holds immense value in minimally invasive endoscopic surgery, enabling enhanced navigation, accurate surface reconstruction, and immersive augmented reality experiences~\cite{collins2020augmented, zhang2020real}. However, due to complex internal environments, low lighting conditions, and sparse feature textures, accurate surgical depth estimation remains a challenging task~\cite{shao2022self}. Traditional multi-view geometry-based methods such as structure from motion (SfM)~\cite{rattanalappaiboon2015fuzzy} and simultaneous localization and mapping (SLAM)~\cite{grasa2013visual} perform poorly in surgical scenarios with low lighting and lack of texture. Deep learning method has been widely proposed for depth estimation in natural environments~\cite {sun2023sc, bhat2022localbins}. Due to security, privacy, and professionalism issues, obtaining large-scale, precise surgical ground truth depth information for supervised training is challenging. Researchers therefore have been focusing on the self-supervised learning (SSL) method where the depths are constrained by the geometric relationship between video frames~\cite{arampatzakis2023monocular, ozyoruk2021endoslam}. Shao \textit{et al.}~\cite{shao2022self} utilize appearance flow to resolve inconsistent lighting problems in endoscopic depth estimation. Yang \textit{et al.}~\cite{yang2024self} designed a lightweight framework combining CNN and Transformers to compress the model parameters effectively.

Recently, foundation models have attracted extremely increasing attention for their amazing performances in various tasks~\cite{kirillov2023segment, oquab2023dinov2, huang2024endo}. Leveraging the large amount of parameters and training data with integrated training methods, foundation models can learn highly generalizable information achieving state-of-the-art performance in multiple downstream tasks involving vision, text, and multi-modal inputs~\cite{cui2024surgical}. However, foundation models may experience significant performance degradation when applied to specific domains such as endoscopic scenes~\cite{wang2023sam}. Training a medical-specific foundational model from scratch presents numerous challenges due to the scarcity of annotated data in the medical domain and the inadequate availability of computational resources. Hence, there has been considerable discourse on the adaptation of existing foundational models to various sub-domains, optimizing the utilization of pre-trained model parameters, and fine-tuning foundation models for specific application scenarios~\cite{wu2023self, chen2023sam}. 

The majority of the current adaptation of foundation models to the medical domain focuses on medical image segmentation and detection instead of regression tasks like depth estimation and annotated prompts are still required for the fine-tuning process~\cite{zhang2023customized}. To this end, we make the effort to explore efficiently adapting foundation models to surgical self-supervised depth estimation where only surgical videos are required for fine-tuning. To be specific, we design Dynamic Vector-Based Low-Rank Adaptation (DV-LoRA) which requires only a small number of parameters to be fine-tuned specifically for medical scenarios. We also design a Convolution Neck block to enhance the model's ability to capture high-frequency information thus facilitating more accurate depth estimation. Meanwhile, to improve the universality of adaptation based on the fact that endoscopic camera intrinsic parameters are not always given, we add a decode head to estimate the camera's intrinsic parameters simultaneously leading our method can be applied with only surgical videos from any unknown cameras.

Our key contributions can be summarized as follows:
\begin{itemize}
    \item[--] We design DV-LoRA and a Convolution Neck block to efficiently adapt foundation models to surgical scene depth estimation with an exceptionally small amount of trainable parameters resulting in low computational resources and short training time.
    
    \item[--] We present a self-supervised adaptation strategy where the depth, ego-motion, and camera's intrinsic parameters estimations are trained in parallel. Our method can be adapted to only surgical videos from any unknown camera which is broadly applicable to most surgical video datasets.
    
    \item[--] Extensive experiments on two publicly available datasets have demonstrated the superior performance of our proposed method over other state-of-the-art SSL depth estimation methods with significantly fewer trainable parameters. It also reveals the important prospects of our proposed model in the endoscopic domain.

\end{itemize}

\section{Method}

\begin{figure}[t]
\centering
\includegraphics[width=1\linewidth]{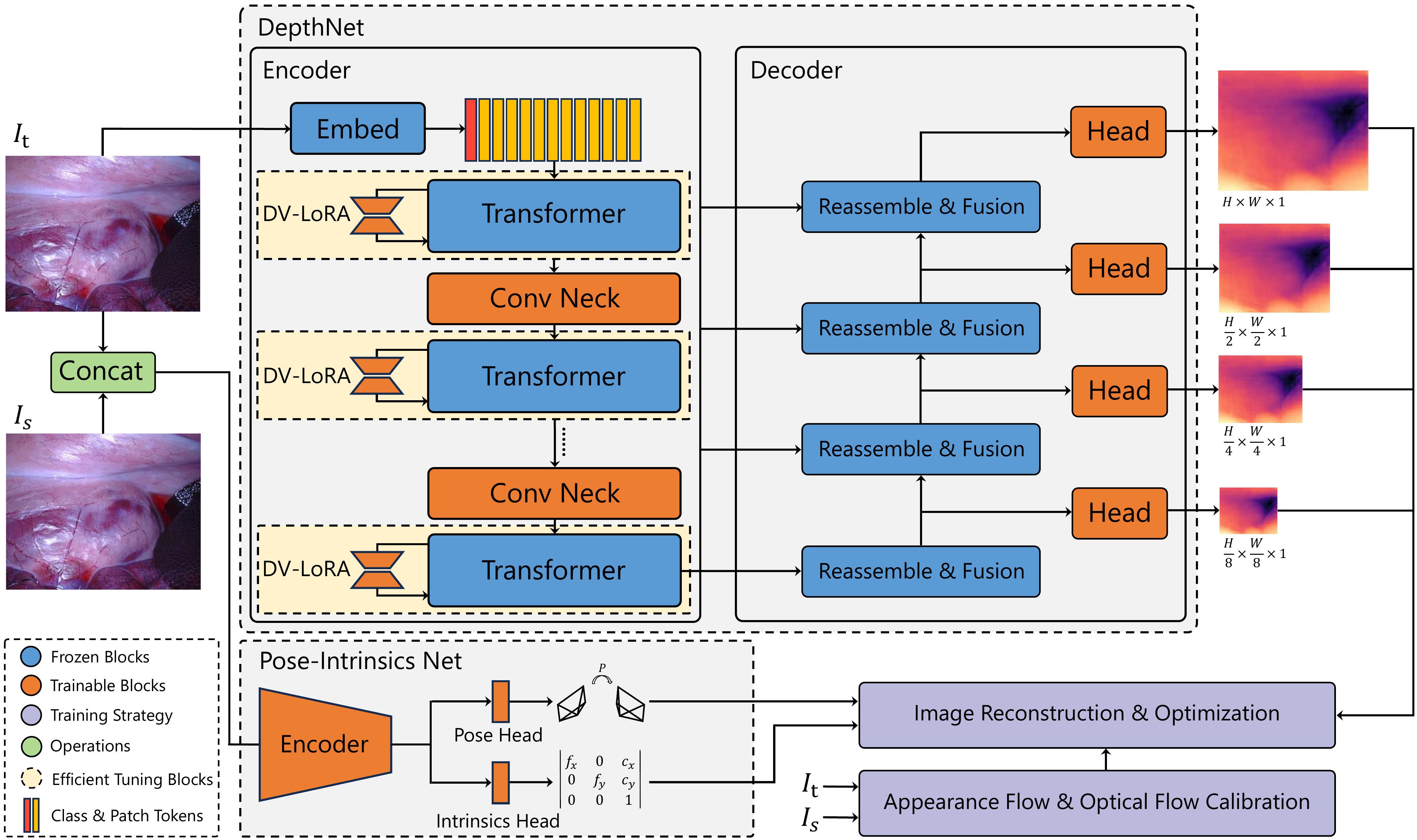}
\caption{Illustration of the proposed Endoscopic Depth Any Camera (EndoDAC) SSL depth estimation framework. ViT-based encoder and DPT-liked decoder pre-trained from Depth Anything~\cite{yang2024depth} are employed for DepthNet. We utilize a small amount of trainable parameters (1.6M) including Dynamic Vector-Based LoRA (DV-LoRA), Convolutional Neck blocks and Multi-Scale Decoders to fine-tune the model. In Pose-Intrinsics Net, ego-motion and camera intrinsic parameters are predicted with the same encoder and separate decoders.}
\label{fig:main}
\end{figure}

\subsection{Preliminaries}

\subsubsection{Foundation Models for Depth} Foundation Models generally refer to powerful pre-trained models trained on extensive amounts of data which enable them to exhibit strong generalization capabilities across multiple tasks and scenarios. Dense Prediction Transformer (DPT)~\cite{ranftl2021vision} is a depth estimation foundation model based on Vision Transformer (ViTs). DINOv2~\cite{oquab2023dinov2} is a semantic foundation model suitable for many vision tasks including depth estimation with separate decode decoders. In this work, we aim to adapt Depth Anything (DA)~\cite{yang2024depth}, which is a depth estimation foundation model trained on large-scale labeled and unlabeled data, to endoscopic scenes.

\subsubsection{Low-Rank Adaptation (LoRA)~\cite{hu2021lora}} LoRA was proposed to fine-tune large-scale foundation models to downstream tasks. It was motivated by the observation that the pre-trained large model's learning ability remains unaffected even when randomly projected onto a smaller subspace. LoRA achieves a significant reduction in the number of trainable parameters for downstream tasks by incorporating trainable rank decomposition matrices into each layer of the Transformer architecture while keeping the pre-trained model weights frozen. To be specific, for a pre-trained weight matrix $W_{0}\in \mathbb{R}^{d \times k}$, LoRA modifies the update to:
\begin{equation}
h = W_{0}x + \Delta Wx = W_{0}x + BAx.
\end{equation}
where $B\in \mathbb{R}^{d \times r}, A\in \mathbb{R}^{r \times k}$ with the rank $r\ll min(d,k)$; $ W_{0}$ is frozen during training and only $A$ and $B$ receive gradient updates.

\subsection{Proposed Framework: EndoDAC }
As illustrated in Fig.~\ref{fig:main}, The architecture of our proposed Endoscopic Depth Any Camera (EndoDAC) framework aims to adapt the depth estimation foundation model - Depth Anything~\cite{yang2024depth} - to the endoscopic domain in a self-supervised manner with minimal training cost. The framework mainly contains two sections: DepthNet and Pose-Intrinsics Net. The DepthNet estimates the multi-scale depth map of a single endoscopic image, while Pose-Intrinsics Net estimates the motion variation between adjacent images and the camera's intrinsic parameters. The DepthNet consists of a ViT-based encoder and a DPT-liked decoder with pre-trained weights. We implement the trainable DV-LoRA layers and Convolutional Neck blocks with the frozen transformer blocks to efficiently fine-tune the model. The Pose-Intrinsics Net estimates the camera ego-motion and intrinsic parameters with the same encoder but separate decoders. The estimated depth is reprojected back to the 2-D plane with the ego-motion information to generate the reconstructed image. The model can therefore be optimized by minimizing the loss between the reconstructed image and the target image. Finally, our EndoDAC only requires endoscopic videos for training and can be applied to any surgical videos without giving the camera intrinsic information.

\subsubsection{DepthNet}

Different from fine-tuning the whole model, EndoDAC freezes the model and adds trainable DV-LoRA layers, Convolutional Neck blocks, and Multi-Scale Decoders, which largely reduces the required memory and computation resources for training and also benefits from convenient deployment.

\begin{figure}[t]
\centering
\includegraphics[width=0.8\linewidth]{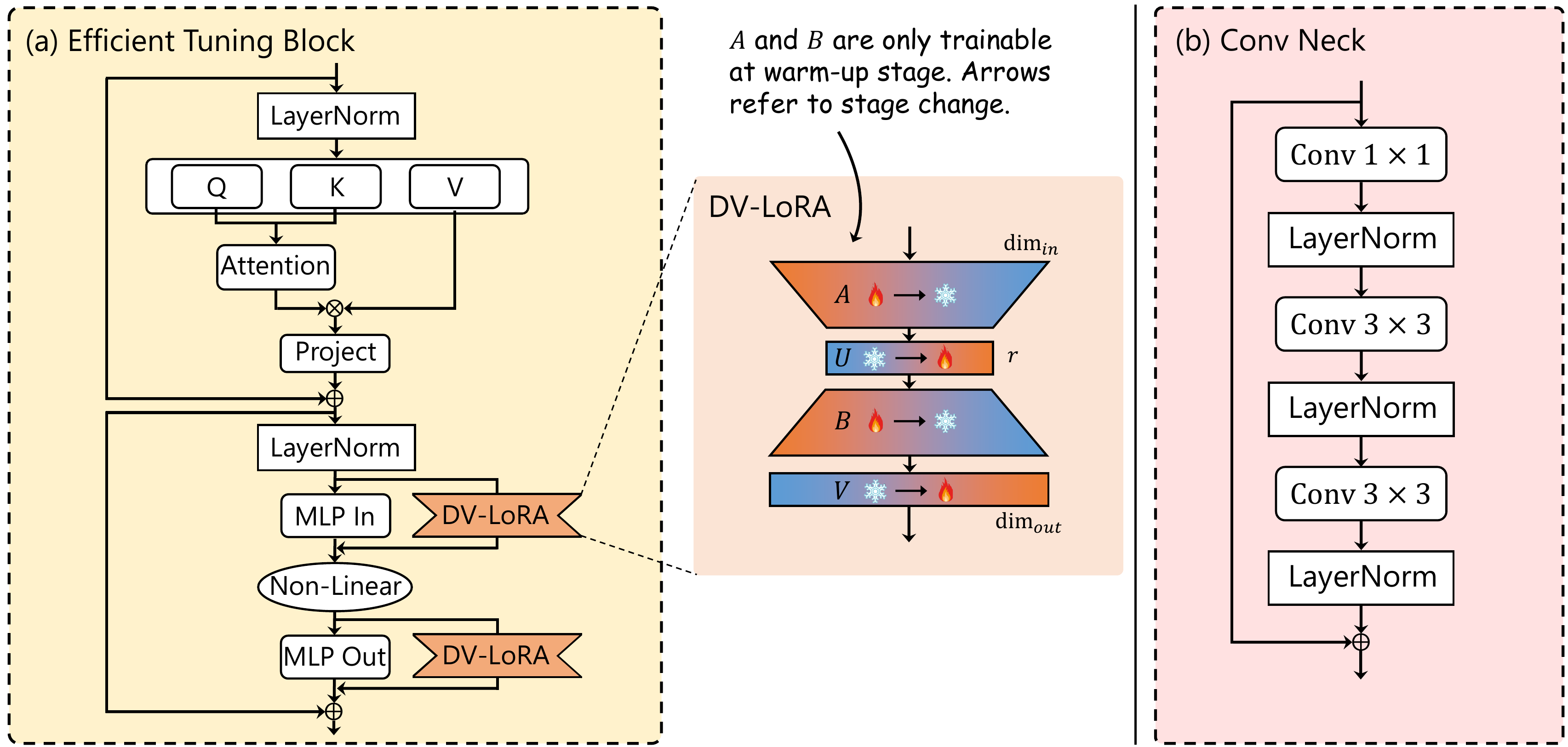}
\caption{Illustration of (a) Transformer Efficient Tuning Block with DV-LoRA and (b) Convolution Neck block. In DV-LoRA, we use the gradient color and arrows to represent the dynamic variation between training and frozen states. }
\label{fig:block}
\end{figure}

A Vision Transformer (ViT), which DA is based on, is used as the backbone of the encoder. Different from conventional LoRA design, we innovatively introduce Dynamic Vector-Based Low-Rank Adaptation (DV-LoRA) to fine-tune the model more efficiently. Fig.~\ref{fig:block}(a) presents our fine-tuning architecture where DV-LoRA is only applied in two MLP layers for more comprehensive adaptation. In contrast to LoRA, our DV-LoRA is expressed as:
\begin{equation}
\begin{aligned}
x_{out} &=\hat{W} x_{in}=W x_{in}+ \Lambda_v B \Lambda_u A x_{in}, \\
\end{aligned}
\end{equation}
where $x_{in}, x_{out}$ are inputs and outputs of MLP layers; $W$ is the frozen projection layer; $A$ and $B$ are trainable LoRA layers; $\Lambda_v$ and $\Lambda_u$ are trainable vectors $U$ and $V$ in diagonal matrices form. At the beginning of training, trainable vectors $U$ and $V$ are frozen while only LoRA layers $A$ and $B$ are trainable. After a warm-up phase, the state of DV-LoRA changes dynamically where $A$ and $B$ are frozen and $U$ and $V$ become trainable. We only train LoRA layers with a good initialization and utilize trainable vectors to fine-tune the proposed model with fewer parameters. 

Furthermore, as demonstrated by~\cite{park2021vision}, ViTs tend to weaken high-frequency signals, which could have negative effects on depth estimation. Therefore, inspired by~\cite{yao2024vitmatte}, we employ Convolution Neck blocks to enhance our method. We incorporate a Convolutional Neck block after each $3^{rd}, 6^{th}, 9^{th}$ and $12^{th}$ transformer efficient tuning block. The details are presented in Fig.~\ref{fig:block}(b) where three convolutional layers with LayerNorm and a residual connection are utilized to feed forward to the results of transformer blocks.

We utilize the DPT-liked decoder to estimate the depth maps. Different from the previous approach where only one depth decoder is applied at the highest resolution, we propose to exploit Multi-Scale Decoders at different fusion levels to generate multi-scale depth maps. Therefore, the feature representations are progressively reassembled and fused at various resolutions. The reassemble and fusion blocks are frozen and only depth decoders are trainable.

\subsubsection{Pose-Intrinsics Net}

Simultaneous learning of depth and motion depends on the relationship between adjacent frames, related depth map and camera matrix:

\begin{equation}
z^{'} p^{'} = KRK^{-1}zp + Kt, 
\label{equation:reproj}
\end{equation}
where K refers to the intrinsic matrix given by 
$
K = \begin{vmatrix}
f_{x} & 0 & x_{0} \\
0 & f_{y} & y_{0} \\
0 & 0 & 1 \\
\end{vmatrix},
\label{equation:K}
$
$p$ and $p^{'}$ are pixel coordinates before and after the transformation of rotation matrix $R$ and translation vector $t$; $z$ and $z^{'}$ are corresponding depths. Previous work~\cite{gordon2019depth} has demonstrated that given Equation.~(\ref{equation:reproj}), no $\tilde{K}$ and $\tilde{R}$ exists such that $\tilde{K}\tilde{R}\tilde{K}^{-1} = KRK^{-1}$ leading the estimation of $K$, $R$ and $t$ converges simultaneously. Therefore, we utilize a shared ResNet~\cite{he2016deep} encoder which takes two adjacent color images as input and outputs the 6-DoF relative pose and camera intrinsics with two separate decoders.

\subsubsection{Self-supervised Depth and Ego-motion Estimation}

With depth $z$, camera intrinsics $K$, rotation matrix $R$ and translation vector $t$ predicted by the networks and given source image $I_{s}$, the reconstructed Image $I_{s\to t}$ is obtained with the re-projection$(\pi )$ defined by Equation.~(\ref{equation:reproj}) described as:
\begin{equation}
I_{s\to t} = \pi \left (  z,K,R,t,I_{s}\right ).
\end{equation}

We utilize an appearance flow network and an optical flow network proposed in~\cite{shao2022self} to compensate for the inconsistent lighting problem.
A photometric loss combing $\mathcal{L}_1$ loss and structural similarities (SSIM)~\cite{wang2004image} is used to assess the image difference defined by:
\begin{equation}
\mathcal{L}_p=\alpha \frac{1-\operatorname{SSIM}\left(I_t, I_{s \rightarrow t}\right)}{2}+(1-\alpha)\left|I_t-I_{s \rightarrow t}\right|.
\end{equation}

An edge-aware loss~\cite{godard2019digging} is also used to maintain the edges defined by:
\begin{equation}
\mathcal{L}_e=\left|\partial_x d\right| e^{-\left|\partial_x \mathbf{I}\right|}+\left|\partial_y d\right| e^{-\left|\partial_y \mathbf{I}\right|},
\end{equation}
where $d$ represents the mean-normalized inverse depth of $I$.

\section{Experiments and Results}


\noindent \textbf{SCARED\footnote{\url{https://endovissub2019-scared.grand-challenge.org/}} Dataset.} SCARED was first proposed for a challenge in MICCAI 2019 containing 35 endoscopic videos with 22950 frames of fresh porcine cadaver abdominal anatomy collected with a da Vinci Xi endoscope. Each video is accompanied by ground truth depth maps collected by a projector and ground truth poses and camera intrinsic. We followed the split scheme in~\cite{shao2022self} where the SCARED dataset is split into 15351, 1705, and 551 frames for the training, validation and test sets, respectively.

\noindent \textbf{Hamlyn\footnote{\url{https://hamlyn.doc.ic.ac.uk/vision/}} Dataset.} Hamlyn is a laparoscopic and endoscopic video dataset taken from various surgical procedures with challenging in vivo scenes. We followed the selection in~\cite{recasens2021endo} with 21 videos for validation.

\noindent \textbf{Implementation Details.} The framework is implemented with PyTorch on NVIDIA RTX 3090 GPU. We utilize two AdamW~\cite{loshchilov2017decoupled} optimizers separately for DepthNet and Pose-Intrinsics Net with initial learning rates of $1 \times 10^{-4}$. The rank for DV-LoRA is set to 4, warm up step is set to 5000 and  $\alpha = 0.85 $ for $\mathcal{L}_p$. We propose the same training augmentations followed~\cite{shao2022self, yang2024self} and batch size is set to 8 with 20 epochs in total. 

\noindent \textbf{Evaluation Settings.} Following~\cite{shao2022self, yang2024self, yang2024depth, cui2024surgical}, we compute the 5 standard metrics: Abs Rel, Sq Rel, RMSE, RMSE log and $\delta$ for evaluation. We re-scale the predicted depth map by a median scaling method~\cite{shao2022self, zhou2017unsupervised, cui2024surgical} during evaluation. We also perform a 5-frame pose evaluation following~\cite{shao2022self} and adopt the metric of absolute trajectory error (ATE). 

\subsection{Results}

\begin{table}[!h]
\caption{Quantitative depth comparison on SCARED dataset of SOTA self-supervised learning depth estimation methods. The best results are in bold and the second-best results are underlined. "G.I." refers to given camera intrinsic parameters. "Total." and "Train." refer to the total and trainable parameters utilized in DepthNet.}
\fontsize{8}{10}\selectfont
\centering
\resizebox{\textwidth}{!}{
\begin{tabular}{c|c|c|c|ccccc|c|c|c}
\toprule 
 & Method & Year & G.I. & Abs Rel $\downarrow$ & Sq Rel $\downarrow$ & RMSE $\downarrow$ & RMSE log $\downarrow$ & $\delta \uparrow$ & Total.(M) & Train.(M) & Speed (ms) \\ \midrule 
\multicolumn{1}{c|}{\multirow{11}{*}{\rotatebox{90}{SCARED}}} & Fang et al.~\cite{fang2020towards} & 2020 & \checkmark & 0.078 & 0.794 & 6.794 & 0.109 & 0.946 & 136.8 & 136.8 & - \\ 
& Defeat-Net~\cite{spencer2020defeat} & 2020 & \checkmark & 0.077 & 0.792 & 6.688 & 0.108 & 0.941 & 14.8 & 14.8 & -  \\ 
& SC-SfMLearner~\cite{bian2019unsupervised} & 2019 & \checkmark & 0.068 & 0.645 & 5.988 & 0.097 & 0.957 & 14.8 &14.8 & -  \\ 
& Monodepth2~\cite{godard2019digging} & 2019 & \checkmark & 0.069 & 0.577 & 5.546 & 0.094 & 0.948 & 14.8 & 14.8 & - \\ 
& Endo-SfM~\cite{ozyoruk2021endoslam} & 2021 & \checkmark & 0.062 & 0.606 & 5.726 & 0.093 & 0.957 & 14.8 & 14.8 & -  \\
& AF-SfMLearner~\cite{shao2022self} & 2022 & \checkmark & 0.059 & 0.435 & 4.925 & 0.082 & 0.974 & 14.8 & 14.8 & 8.0 \\ 
& Yang et al.~\cite{yang2024self} & 2024 & \checkmark & 0.062 & 0.558 & 5.585 & 0.090 & 0.962 & 2.0 & 2.0 & - \\ 
& DA (zero-shot)~\cite{yang2024depth} & 2024 & \checkmark & 0.084 & 0.847 & 6.711 & 0.110 & 0.930 & 97.5 & - & 13.8 \\ 
&  DA (fine-tuned)~\cite{yang2024depth} & 2024 & \checkmark & 0.058 & 0.451 & 5.058 & 0.081 & 0.974 & 97.5 & 11.2 & 13.8\\ \cline{2-12}
& \textbf{EndoDAC (Ours)} & - & \checkmark & \textbf{0.051} & \textbf{0.341} &  \textbf{4.347} & \textbf{0.072} & \textbf{0.981} & 99.0 & 1.6 & 17.7 \\ 
& \textbf{EndoDAC (Ours)} & - & \xmark & \underline{0.052} & \underline{0.362} & \underline{4.464} & \underline{0.073} & \underline{0.979} & 99.0 & 1.6 & 17.7 \\ \midrule 
\multicolumn{1}{c|}{\multirow{5}{*}{\rotatebox{90}{Hamlyn}}} & Endo Depth \& Motion~\cite{recasens2021endo} & 2021 & \checkmark & 0.185 & 5.424 & 16.100 & 0.225 & 0.732 & - & - & - \\ 
& AF-SfMLearner~\cite{shao2022self} & 2022 & \checkmark & 0.168 & 4.440 & 13.870 & 0.204 & 0.770 & 14.8 & 14.8 & 7.7 \\ 
& DA (fine-tuned)~\cite{yang2024depth} & 2024 & \checkmark & 0.170 & 4.413 & 13.920 & 0.205 & 0.765 & 97.5 & - & 12.5 \\ \cline{2-12}
& \textbf{EndoDAC (Ours)} & - & \checkmark & \textbf{0.138} & \textbf{2.796} & \textbf{11.491} & \textbf{0.171} & \textbf{0.813}  & 99.0 & 1.6 & 15.7 \\ 
& \textbf{EndoDAC (Ours)} & - & \xmark & \underline{0.156} & \underline{3.854} & \underline{12.936} & \underline{0.193} & \underline{0.791} & 99.0 & 1.6 & 15.7 \\ \bottomrule 
\end{tabular}} 
\label{tab:main}
\end{table}

\begin{figure}[t]
\centering
\includegraphics[width=0.9\linewidth]{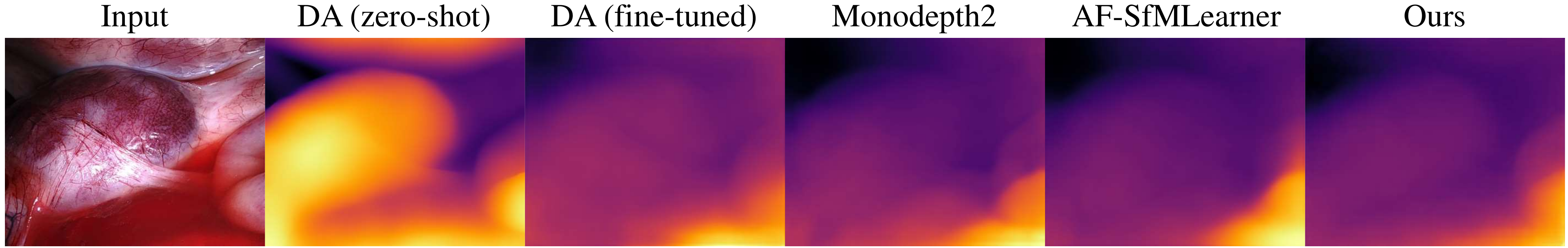}
\caption{Qualitative depth comparison on the SCARED dataset. }
\label{fig:vis_1}
\end{figure}

\noindent \textbf{Depth Estimation.} The proposed method is compared with several SOTA self-supervised methods~\cite{fang2020towards, spencer2020defeat, bian2019unsupervised, godard2019digging, ozyoruk2021endoslam, shao2022self, yang2024self, recasens2021endo}, Depth Anything ~\cite{yang2024depth} model with and without full parameter fine-tuning on two datasets. Note that we zero-shot evaluated on Hamlyn with the models trained on SCARED. All the other compared methods involving training have a total epoch greater or equal to 40 while our approach only trains for 20 epochs. Table~\ref{tab:main} presents the comparison of quantitative results of the aforementioned methods. Our method exceeds all of the compared methods by a significant margin regardless of the awareness of camera intrinsics. It is also notable that only 1.6 million parameters are trainable for our model which accounts for 1.6\% of the total parameters. Foundation models generally utilize large models leading our framework slower in inference speed compared to other methods, but a speed of 17.7 ms is still capable of real-time implementation which makes our framework applicable for a variety of real-time surgical applications. Fig.~\ref{fig:vis_1} shows several qualitative results visualization. We can observe that our method generates a more accurate geometry relation within the depth map while also preserving the global smoothness of tissues. More visualization of depth and 3D reconstruction are presented in Supplementary. 

\noindent \textbf{Ablation Studies.} To further demonstrate the validity of our proposed model, we conduct ablation studies on the different modules of EndoDAC. As presented in Table~\ref{tab:ablation}, the ablation studies demonstrate the effectiveness of each module.

\begin{table}[!h]
\caption{Ablation study on the modules of EndoDAC with estimated intrinsics. Specifically, we (i) use the original LoRA~\cite{hu2021lora} to replace DV-LoRA; (ii) remove the Convolution Neck blocks; (iii) use a single depth decoder to replace the multi-scale decoders.}
\fontsize{8}{10}\selectfont
\centering
\resizebox{0.8\textwidth}{!}{
\begin{tabular}{c|c|c|ccccc}
\hline 
DV-LoRA & Conv Neck & M.S. Decoders & Abs Rel $\downarrow$ & Sq Rel $\downarrow$ & RMSE $\downarrow$ & RMSE log $\downarrow$ & $\delta \uparrow$ \\ \hline 
\checkmark & \xmark & \xmark & 0.054 & 0.397 & 4.718 & 0.076 & 0.978\\ 
\xmark & \checkmark & \xmark & 0.053 & 0.408 & 4.768 & 0.076 & 0.977 \\  
\xmark & \xmark & \checkmark & 0.051 & 0.379 & 4.621 & 0.074 & 0.977  \\ 
\checkmark & \checkmark & \xmark & 0.052 &0.380& 4.598& 0.075 & 0.977  \\  
\checkmark & \xmark & \checkmark & 0.052 & 0.378 & 4.614 & 0.075 & 0.978   \\ 
\xmark & \checkmark & \checkmark & 0.053 & 0.366 & 4.516 & 0.074 & 0.978 \\  
\checkmark & \checkmark & \checkmark & 0.052 & 0.362 & 4.464 & 0.073 & 0.979   \\ \hline 

\end{tabular}} 
\label{tab:ablation}
\end{table}

\noindent \textbf{Pose and Intrinsics Estimation.} 
Two sequences are selected followed~\cite{yang2024depth, shao2022self} for evaluation of pose and intrinsic estimation. Pose estimation is evaluated on two sequences separately while intrinsic estimation is evaluated with a weighted average percentage error on two sequences. The results are presented in Table~\ref{tab:pose} and Table~\ref{tab:intrinsic}. Table~\ref{tab:pose} shows our proposed method obtains satisfactory performances on pose estimation with or without given intrinsic parameters. Our proposed method can also estimate accurate camera intrinsic parameters with a maximum percentage error of 4.02\%. Visualization of pose estimation trajectories of two sequences are presented in Supplementary. 

\begin{table*}
\begin{floatrow}
\capbtabbox{
\resizebox{0.5\textwidth}{!}{
\begin{tabular}{c|c|c|c}
                    \toprule  
                    Method & G.I. & ATE $\downarrow$ (Seq.1) & ATE $\downarrow$ (Seq.2) \\ \midrule 
                    Monodepth2~\cite{godard2019digging} & \checkmark & 0.0769 & 0.0554 \\
                    Endo-SfM~\cite{ozyoruk2021endoslam} & \checkmark & 0.0759 & 0.0500 \\ 
                    AF-SfMLearner~\cite{shao2022self} & \checkmark & 0.0742 & 0.0478 \\ \midrule 
                    \textbf{EndoDAC (Ours)} & \checkmark & 0.0741 & 0.0512 \\ 
                    \textbf{EndoDAC (Ours)} & \xmark & 0.0762 & 0.0487 \\ \bottomrule
                \end{tabular}}
}{
 \caption{Quantitative pose estimation comparison on two selected sequences. "G.I." refers to given camera intrinsic parameters.}
 \label{tab:pose}
}
\capbtabbox{
\resizebox{0.39\textwidth}{!}{
                \begin{tabular}{c|c|c}
                    \toprule 
                     Intrinsics & Ground Truth & Percentage Error $\downarrow$(\%) \\ \midrule
                     fx & 0.82 & 3.17 \\ \midrule
                     fy & 1.02 & 4.02 \\ \midrule
                     cx & 0.5 & 0.70 \\ \midrule
                     cy & 0.5 & 1.70 \\ \bottomrule
            
                \end{tabular}}
}{
 \caption{Quantitative intrinsic estimation results.}
 \label{tab:intrinsic}
 \small
}
\end{floatrow}
\end{table*}

\section{Conclusion}
In this paper, we propose an efficient framework to adapt foundation models to SSL endoscopic depth estimation. Different from previous methods, we introduce DV-LoRA and Convolutional Neck to fine-tune depth foundation models to the endoscopic domain with limited trainable parameters and training costs. We also utilize a separate decoder to estimate camera intrinsic and ego-motion simultaneously. Our method is widely applicable to surgical datasets where only surgical videos are required for training. Extensive experiments demonstrate the superior performance of our proposed method even without knowledge of the camera and intrinsic parameters information.

\subsection*{Acknowledgements}
This work was supported by Hong Kong Research Grants Council (RGC) Collaborative Research Fund (C4026-21G), General Research Fund (GRF 14211420 \& 14203323),  Shenzhen-Hong Kong-Macau Technology Research Programme (Type C) STIC Grant SGDX20210823103535014 (202108233000303).

\bibliography{mybib}{}
\bibliographystyle{splncs04}

\newpage
\section*{Supplementary Materials for ``EndoDAC: Efficient Adapting Foundation Model for Self-Supervised Depth Estimation from Any Endoscopic Camera''}

\begin{table}[!h]
\caption{Definition of evaluation metrics.}
\fontsize{8}{10}\selectfont
\centering
\resizebox{0.6\textwidth}{!}{
\begin{tabular}{c|c}
\toprule 
Metrics Name & Definition \\ \midrule 
$AbsRel$ & $\frac{1}{\left| \textbf{D}\right|}\sum_{d\in \textbf{D}}^{}\left|d^{*}-d \right|/d^{*}$ \\ \midrule 
$SqRel$ & $\frac{1}{\left| \textbf{D}\right|}\sum_{d\in \textbf{D}}^{}\left|d^{*}-d \right|^{2}/d^{*}$ \\ \midrule 
$RMSE$ & $\sqrt{\frac{1}{\left| \textbf{D}\right|}\sum_{d\in \textbf{D}}\left| d^{*}- d\right|^{2}}$ \\ \midrule 
$RMSElog$ & $\sqrt{\frac{1}{\left| \textbf{D}\right|}\sum_{d\in \textbf{D}}\left| logd^{*}- logd\right|^{2}}$ \\ \midrule 
$\delta$ & $\frac{1}{\left| \textbf{D}\right|}\left| \left\{ d\in \textbf{D}|max(\frac{d^{*}}{d}, \frac{d}{d^{*}} < 1.25) \right\} \right| \times 100\%$  \\ \bottomrule 
\end{tabular}} 
\label{tab:sup}
\end{table}

\begin{figure}[th]
\centering
\includegraphics[width=0.9\linewidth]{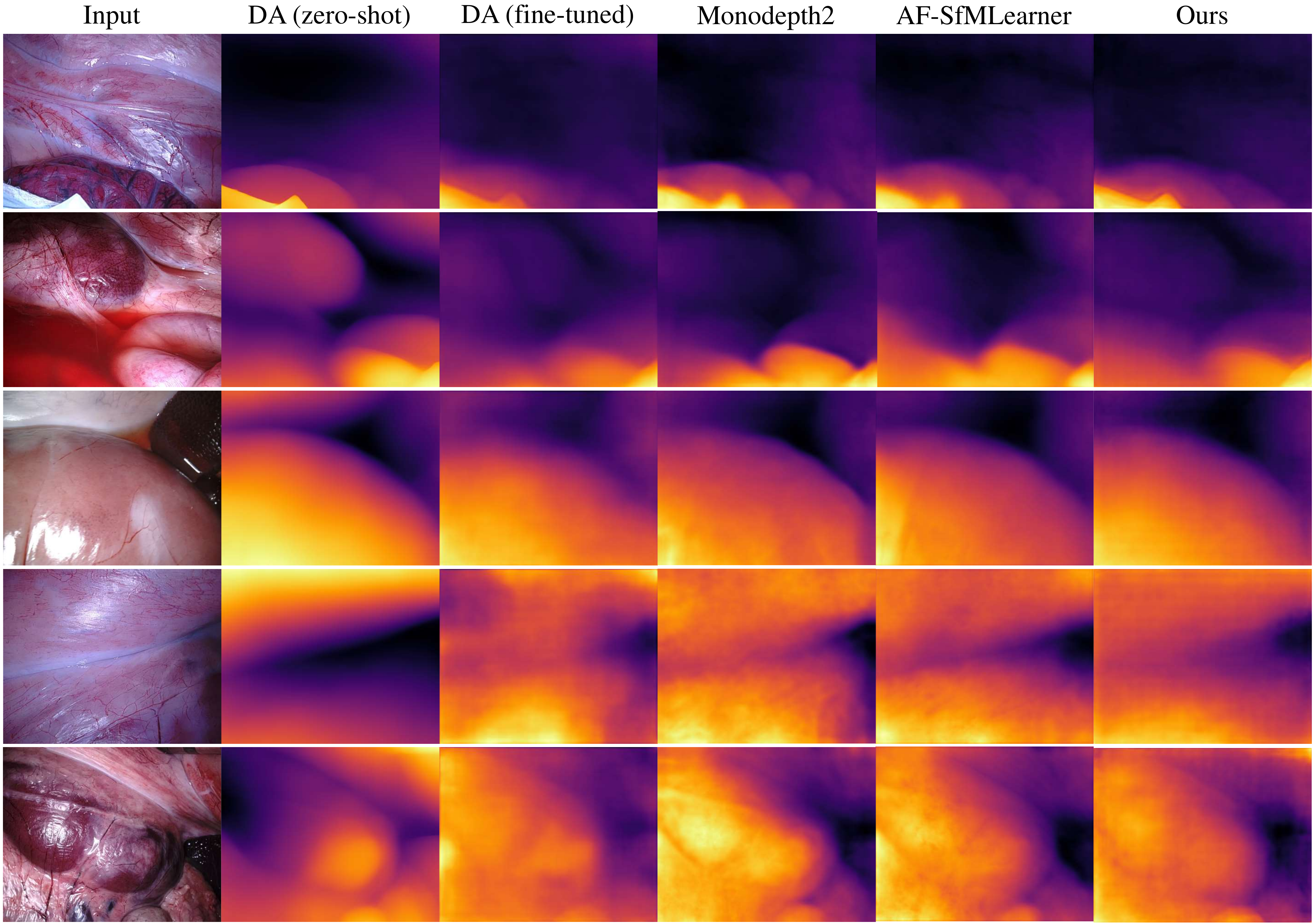}
\caption{Qualitative depth comparison on the SCARED dataset. }
\label{fig:sup_vis_1}
\end{figure}

\begin{figure}[th]
\centering
\includegraphics[width=1\linewidth]{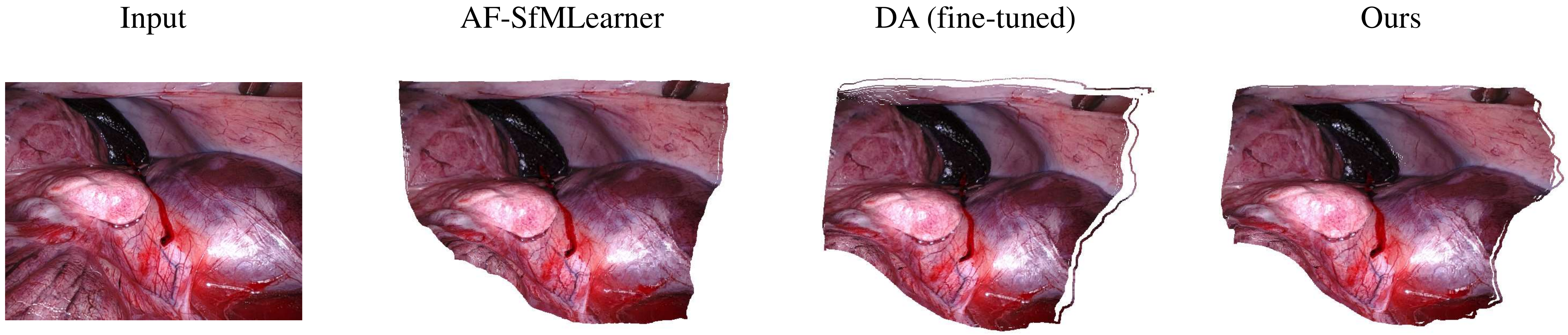}
\caption{Qualitative 3D reconstruction comparison on the SCARED dataset. }
\label{fig:sup_vis_2}
\end{figure}

\begin{figure}[th]
\centering
\includegraphics[width=1\linewidth]{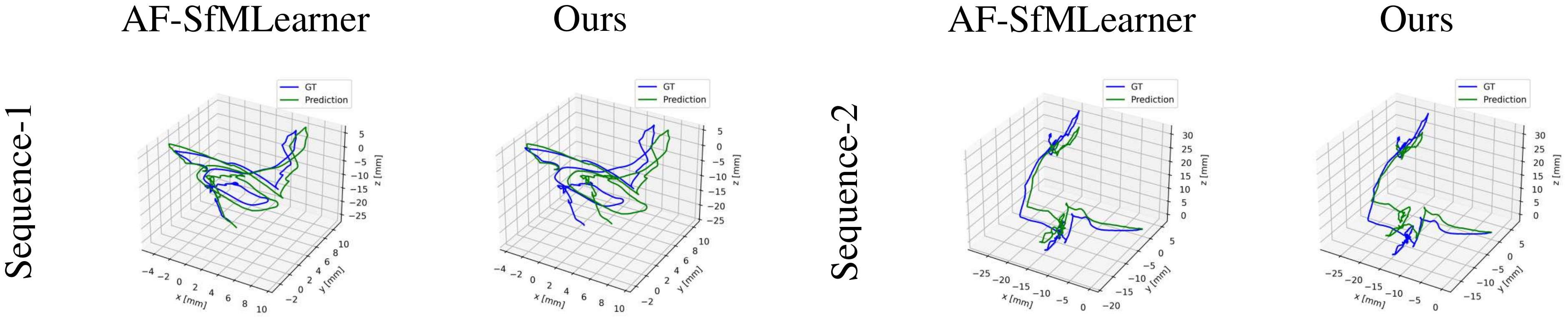}
\caption{Qualitative pose estimation comparison on the SCARED dataset. }
\label{fig:sup_vis_3}
\end{figure}

\end{document}